%% file: hdi2014tochi.tex
\documentclass{hcj}

\title{Human-Data Interaction: \\
  The Human Face of the Data-Driven Society}
  
\author{
Richard Mortier\affil{University of Nottingham}
        \and Hamed Haddadi\affil{Queen Mary University of London}
        \and Tristan Henderson\affil{University of St. Andrews}
        \and Derek McAuley\affil{University of Nottingham}
        \and Jon Crowcroft\affil{University of Cambridge}
        }
        
\authorrunning{Mortier, Haddadi, Henderson, McAuley and Crowcroft}

\begin{document}

\maketitle

\begin{abstract}
\input{abstract}

\end{abstract}

%
%
%

\input{intro}
\input{hdi}
\input{legibility}
\input{agency}
\input{negotiability}

\input{conclusion}

%




\bibliography{hdi}

\end{document}

%% file: abstract.tex
The increasing generation and collection of personal data has created
a complex ecosystem, often collaborative but sometimes combative,
around companies and individuals engaging in the use of these data. We
propose that the interactions between these agents warrants a new
topic of study: Human-Data Interaction (HDI). In this paper we discuss
how HDI sits at the intersection of various disciplines, including
computer science, statistics, sociology, psychology and behavioural
economics. We expose the challenges that HDI raises, organised into
three core themes of legibility, agency and negotiability, and we present
the HDI agenda to open up a dialogue amongst interested parties in the
personal and big data ecosystems.





%% file: intro.tex
\section{Introduction}
\label{sec:intro}

The process of moving from a world where computing is siloed and specialised, to a world where computing is ubiquitous and everyday, continues. In many parts of the world, networked computing is now mundane, both as a foreground technology (e.g.,~smartphones, tablets) and in the background (e.g.,~road traffic management, financial systems). This has permitted, and continues to permit, new gloss on existing interactions as well as fundamentally new interactions (e.g.,~online banking, massively scalable distributed real-time gaming).

We observe that human-computer interaction (HCI) research has traditionally focused on the interactions between humans and computers-as-artefacts, i.e.,~devices to be interacted with. As described by Grudin~\cite{grudin90:inter,grudin90:comput.reach}, the focus of work in HCI varies from psychology~\cite{card:psychology} hardware to software to interface, and subsequently deeper into the organisation. This trend, of the focus to move outward from the relative simple view of an operator using a piece of hardware, continued with e.g.,~Bowers and Rodden~\cite{bowers93:explod.inter} considering the richness of the inter-relationships between users and computer systems as those systems have pervaded organisations and become networked, and thus the need to ``explode the interface''. However in this paper we will not focus on the HCI history and the wide spectrum of research interests and approaches.

We believe that a broad range of existing work in many areas points to a need to emphasise another facet of the very general topic of how people interact with computer systems. Specifically, the continuing and accelerating trend towards truly ubiquitous and pervasive computing is beginning to raise the issue of how people should interact with {\bf data}. A complex ecosystem, often collaborative but sometimes combative~\cite{browneconomics}, is forming around companies and individuals engaging in the use of these data. We propose placing the \emph{human} at the centre of the flows of data, and providing mechanisms for citizens to interact with these systems and data \emph{explicitly}: {\bf Human-Data Interaction (HDI)}.

Others have previously coined this phrase in relation to interfaces enabling interaction by individuals with specific datasets~\cite{embodiedHDI,Cafaro:2012:UEA:2370216.2370309}. We believe that a broader conception of HDI is required, and have previously stated briefly some of the other challenges and opportunities that we perceive in HDI~\cite{mortierchallenges}. Now, informed by input from perspectives ranging from computer systems to law to sociology to economics, we survey the influences and factors that led us to this belief, presenting the following contributions in this paper:

\begin{itemize}
\item We elaborate our conception of HDI as it arises from various motivating factors, relating it specifically to existing work and specific prior uses of the term~(\S\ref{s:hdi}); and

\item We expose the challenges that we believe are raised by HDI, organised into three core themes: \emph{legibility}~(\S\ref{s:legibility}), \emph{agency}~(\S\ref{s:agency}), and \emph{negotiability}~(\S\ref{s:negotiability}).
\end{itemize}

Our goal in this paper is to open up the dialogue amongst parties interested in making the \emph{human} explicit in the data ecosystem.

%% file: hdi.tex
\section{Defining Human-Data Interaction}
\label{s:hdi}

One common effect of the increasing pervasiveness of electronic computation in our environments and our lives is that \emph{data} are also now ubiquitous -- society is becoming ``data-driven''~\cite{pentland:data-driven-society}. Many of the devices we use (e.g.,~phones, computers), the networks through which they connect (predominantly the Internet, but also alternative technologies such as the fixed and cellular telephone networks), and the interactions we experience that use these technologies (e.g.,~use of credit cards, driving on public highways, online shopping) generate considerable trails of data.

An important distinction seems to be whether data are (consciously) created \emph{by} us (Volunteered data like our Online Social Network profile, or by observed data such as online shopping behaviour~\cite{forum11:_person_data}), or they are inferred and created \emph{about} us by others (this includes machines, programs, as well as people). Although in many cases the release of data through interaction with individual computing systems is relatively conscious (e.g.,~the use of social media such as Facebook, or cloud email services such as GMail), in many others it is considerably less, so particularly where multiple computing systems are involved in combination (e.g.,~the way that credit card transactions or closed-circuit television and automatic number plate recognition creating a personal location history).

These data are accumulated about us by many different organisations; some competing, some collaborating, and they are processed using increasingly sophisticated algorithms to measure and infer increasingly sensitive features of our lives (e.g.,~purchasing behaviour revealing our political or sexual preferences, our state of mind, or our likely future behaviour). This increasing practice of accumulation of data and the increasing importance of these data and inferences drawn from them for our everyday lives drives the need for the study of HDI.

\subsection{The Evolution of Data}

The term HDI has arisen previously in reference to, e.g.,~``the human manipulation, analysis, and sense-making of large, unstructured, and complex datasets''~\cite{embodiedHDI} and ``technologies that use embodied interaction to facilitate the users' exploration of rich datasets''~\cite{Cafaro:2012:UEA:2370216.2370309}. While we do not dispute that the ability to interact directly with rich datasets is important, we believe that the richness of conceptions of data lead to a broader definition of HDI. The dictionary definition of ``data'' is very general and quite mundane~\cite{oed}:

\begin{quote}
  \begin{enumerate}[1.]
  \item As a count noun: an item of information; a datum; a set of data.
  \item As a mass noun.
    \begin{enumerate}[(a)]
    \item Related items of (chiefly numerical) information considered collectively, typically obtained by scientific work and used for reference, analysis, or calculation.
    \item \emph{Computing}. Quantities, characters, or symbols on which operations are performed by a computer, considered collectively. Also (in non-technical contexts): information in digital form.
    \end{enumerate}
  \end{enumerate}
\end{quote}

When compounded with other nouns however, it becomes more interesting:

\begin{quotation}
{\bf data trail} {\em n}. an electronic record of the transactions or activities of a particular person, organisation, etc. Now esp. with reference to a person's financial transactions, telephone and Internet usage, etc.

{\bf data smog} {\em n}. a confusing mass of information, esp. from the Internet, in which the erroneous, trivial, or irrelevant cannot be easily or efficiently separated from what is of genuine value or interest (often in figurative context); obfuscation generated by this; cf. information overload.

{\bf big data} {\em n}. Computing (also with capital initials) data of a very large size, typically to the extent that its manipulation and management present significant logistical challenges; (also) the branch of computing involving such data.
\end{quotation}

In many ways, it is the interplay between the last three definitions that we believe gives rise to the need for a broader conception of HDI: when the data trails of individual, private behaviour are coalesced and analysed as big data; and where the results of that analysis, whether or not correct, are fed back into the data associated with an individual. Data, particularly personal data, can be seen as a \emph{boundary object}~\cite{star89:instit.ecolog.trans.bound.objec,leigh10:this.not.bound.objec}, reflected in the many ways different communities refer to and think of data.

For example, to contrast with big data we see data trails referred to as \emph{small data}~\cite{estrin:universe} where ``$N$ = me'', pertaining to each of us as individuals. We see yet other terms used in other fields: \emph{participatory data}~\cite{shilton:personal} in health, \emph{microdata}~\cite{kum:population-informatics} in population informatics, and \emph{digital footprint}~\cite{madden07:digit.footp} in the digital economy.

Some legal frameworks give quite precise definitions. For example, the EU Directive 95/46/EC states ``{\em\,`personal data' shall mean any information relating to an identified or identifiable natural person (`data subject'); an identifiable person is one who can be identified, directly or indirectly, in particular by reference to an identification number or to one or more factors specific to his physical, physiological, mental, economic, cultural or social identity;}''.\footnote{\url{http://eur-lex.europa.eu/legal-content/en/ALL/?uri=CELEX:31995L0046}} It is this definition that is used to trigger application of data protection law, but the interpretation of this definition varies between national courts and over time. For example, in the past UK courts have often interpreted it more narrowly than those in continental Europe, though more recently the UK Court of Appeal has reinterpreted it, broadening the definition.\footnote{See, e.g.,~\url{http://www.scl.org/site.aspx?i=ed35678}.}

\emph{Open data} has a less formal but also fairly organised definition, accepted by a number of organisations active in that space: ``A piece of data or content is open if anyone is free to use, reuse, and redistribute it --- subject only, at most, to the requirement to attribute and/or share-alike.''~\cite{opendefinition}. Even here, however, nuances creep as this definition of open data may apply differently to particular types of data in various fields, e.g.,~open access (publishing), open source (computer programming), open content (creative industries), or public data (many government data sources), or apply selectively, e.g., distinguishing between open science~\cite{woelfle:open} whereby processes may be shared, or open scientific data which may not be appropriate for all types of research data~\cite{heeney:sharing}.

\subsection{Embracing Human-Data Interaction}

We have argued that data, and particularly personal and open data, are sufficiently rich concepts that HDI should be broadened to encompass more than simply means to interact directly with data. Five specific ways in which we do so are:

\begin{itemize}
\item We refer not just to embodied interaction but the many different forms interaction around data can take. For example, the ways in which we may choose to permit or deny access by third-parties to our personal data.

\item We refer not just to the foreground activity of exploring data, but to background and ambient ways in which we understand data. For example, the notion of periodically receiving a ``data statement'' indicating by whom our data have been accessed and how they have been used.

\item We refer not just to systems for interaction with data directly, but also to the need for systems that enable interaction with the systems that process our data. For example, the need to understand the inferences that might be drawn from our data and thus the consequences of our actions in making personal data available.

\item We do not consider data to be static but dynamic, under constant revision and extension. Specifically, the algorithms and processes that consume data often emit other data as outputs that are then fed in to other processes as inputs of equal validity to raw ``observed'' data.

\item We do not consider simply the individual's relationship to some data, whether by or about them. Data can be seen as boundary object: open to multiple interpretations and the concern of many stakeholders: the individual or group whom the data concerns; those who collected the data; those legally responsible for the data; and the potentially many parties who use, or wish to use, the data.
\end{itemize}

\begin{figure}
  \centering
  \includegraphics[width=0.8\columnwidth]{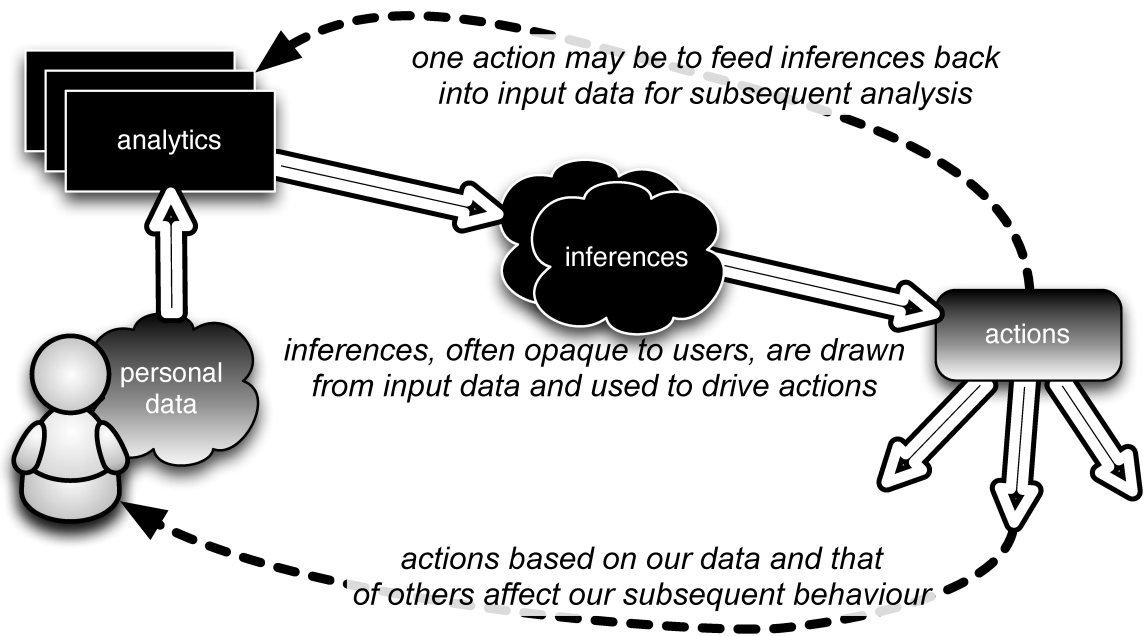}
  \caption{\label{hdi}Human-Data Interaction. Our \emph{personal data} feeds black-box \emph{analytics} algorithms. These output \emph{inferences} driving \emph{actions} whose effects may or may not be visible to us, and which may include changes in our behaviour and the data generated about and by us subsequently.}
\end{figure}

We characterise the key elements of HDI in Figure~\ref{hdi}. Analytics are provided as a ``black box'' (or rather, set of black boxes) within which collated input data are processed in large centralised facilities (data centres). The inferences output by this processing then cause actions, which may include feeding inferences into subsequent analysis by others. We thus identify three core themes by which we organise the remainder of this presentation:

\begin{itemize}
\item \emph{Legibility}, concerned with making data and analytics algorithms both transparent and comprehensible to the people the data and processing concerns. What data are they consuming? What methods are they using to draw inferences? This is often in direct conflict with the fact that these algorithms represent core intellectual property of the companies that implement and run them, and so cannot easily be made public.

\item \emph{Agency}, concerned with giving people the capacity to act within these data systems, to opt-in or to opt-out, to control, inform and correct data and inferences, and so on. That is not to say that we believe \emph{all} users must or will choose to continually exercise this capacity, engaging in detailed, ongoing control and management of their data; but some enthusiasts may choose to, and many people may wish to from time-to-time whether triggered by idle curiosity or in response to perceived potential or realised harm.

\item \emph{Negotiability}, concerned with the many dynamic relationships that arise around data and data processing. This theme encompasses, for example, how individual understanding and attitudes change over time, how society responds to these problems through the formation of social norms around data and data processing, the different legal and regulatory frameworks that may apply in different jurisdictions to different types of data and types of data processing, and the ways in which data is often ambiguous in its subject and its meaning.
\end{itemize}

Although there are certainly other organisations of this complex landscape, e.g.,~\cite{richards:paradoxes}, we believe that the three above -- legibility, agency, negotiability -- provide an effective means to structure discussion of the myriad issues surrounding HDI.

%% file: legibility.tex
\section{Legibility}
\label{s:legibility}

At present, our interactions with online data systems are often opaque to us: there are few online analogues of physical world artefacts such as CCTV signage.\footnote{E.g.,~\url{http://ico.org.uk/for_the_public/topic_specific_guides/cctv}} We argue that it is not enough simply to make these processes transparent: they are often technical, complex and the implications of the data collected and processed are incomprehensible. Rather, we believe that they must be made \emph{legible}, able to be understood by the people they concern. This is a precursor to their ability to consciously exercise agency in situations where personal data being collected and processed. Already recognised in specific contexts, such as consent and withdrawal~\cite{coles-kemp:privacy,literatin,custers:consent}, the need for data to be more legible is becoming pervasive as society becomes more data-driven.

Data created about us are often less well-understood by the subject. For instance, third-party website tracking, when combined with recommender systems and data-mining algorithms~\cite{Krishnamurthy,Butkiewicz:webcomplexity}, can create new data from inferences, such as advertising preferences~\cite{narseoIMC}. Credit-scoring companies such as Experian~\cite{experian} and ``customer science'' companies such as dunnhumby~\cite{dunnhumby} collect and mine shopping and transaction data to both predict and enable behaviours. Not all such data uses are strictly commercial. For instance, personal data can be used to generate data for new crowdsourced applications such as traffic reports on Waze~\cite{waze}, optimised bus routes~\cite{berlingerio:allaboard}, or even mapping informal transport routes that have otherwise never been officially recorded~\cite{digitalmatatus}. But new tools for informing people about their data, and the practices used around these data, are essential~\cite{vandenberg:processing}.

Data created by us arise from our interaction with numerous sensors and technologies. Obvious sources include such recently mundane technologies as Facebook, Twitter and other online social networks (OSNs) and websites more generally~\cite{Krishnamurthy:footprint}. The richness and variety of such data, however, is continually increasingly, particularly with the growing interest in lifelogging and the ``Quantified Self''~\cite{quantself,rooksby:tracking,choe:quantified-selfers,kobsa:personalization}. For example, devices and sensors with which we explicitly interact when monitoring our health (e.g.,~continuous blood glucose monitoring, smart asthma inhalers, bathroom scales that track our weight, or smartphone apps that monitor our sleep patterns). Such devices can create ``people-centric'' sensor trails~\cite{campbell:people-centric}. Related advances in portable medical sensors, affordable personal genomics screening, and other tools for mobile health diagnosis will generate new personal medical datasets~\cite{kumar:transdisciplinary}.

Legibility entails several features. First, we need to become aware that data is being collected, relatively straightforward to achieve as with, e.g.,~recent European legislation requiring websites make clear to users when they use browser cookies. The second, and more complex requirement is that we become aware of the data themselves and their implications. A data-centric view of the world requires that we pay attention to the correctness (in an objective knowledge sense) of data. However, when taking a human-centric view of a data-driven world, it is important to be aware that individuals' viewpoints of data may differ but be valid. Similarly, interpretations of data may vary significantly over time, hence (for example) the recent EU ``right-to-be-forgotten'' empowers individuals to request that search engines remove links to data about them when the information via those links becomes ``inaccurate, inadequate, irrelevant or excessive for the purposes of data processing'', to protect their privacy (though this right is not absolute but is balanced against other fundamental rights.

Simply providing visualisations of data is a starting point, and a well-studied topic within HCI. However, even this can pose problems due to the scale of data involved as Quantified Self app developers have found when presenting the large, detailed, rich data collected about aspects of a single individual, from physical activity to sleep patterns and diet~\cite{chi_vis_chal}. Similar problems arise with data that are inherent ambiguous such as those collected about communities through Internet-of-Things technologies~\cite{Zaslavsky}. However, the potential for data visualisation to reveal aspects of the incentive models associated with the processing of data, and even the details of the processing algorithms themselves, may present more problematic challenges in a commercial environment.

One possible avenue is to engage with artists in attempting to make these very abstract concepts (data, algorithm, inference) legible to users. Early attempts, e.g.,~Tangible Souvenirs\footnote{\url{http://proboscis.org.uk/tag/tangible-souvenirs/}} or Sweat Atoms~\cite{sweat-atoms}, have proved promising. Indeed, it is in this context, specifically Embodied Interactions, that we see first use of the term ``Human-Data Interaction'' of which we are aware~\cite{embodiedHDI,Cafaro:2012:UEA:2370216.2370309}. In his epilogue reflecting on the original book on embodied interaction, Dourish clarifies that tangible computing ``by no means defines or sets the boundaries of embodied interaction or embodied analysis''~\cite{dourish:epilogue}, and extending this route of enquiry to interaction with data may well prove fruitful.

%% file: agency.tex
\section{Agency}
\label{s:agency}

Empowering us to become aware of the fact and implications of collection of our personal data is a beneficial first step. However, putting people at the heart of these data processing systems requires more: we require \emph{agency}, the capacity to act for ourselves within these systems.

Recently, the right to be informed when personal data are collected, is enshrined in legislation such as the European Data Protection Directive. The \textit{right to be forgotten} has also been recently enforced by the European Union Court of Justice.\footnote{An Internet search engine operator is responsible for the processing that it carries out of personal data which appear on web pages published by third parties (retrieved 13 May 2014) \url{http://curia.europa.eu/jcms/upload/docs/application/pdf/2014-05/cp140070en.pdf}} But as the intimacy, ubiquity and importance of the personal data collected about us grows, we require a broader ability to engage with its collection, storage and use, to understand and modify raw data and the inferences drawn from it.

This is more than simply the ability to provide informed consent, though even that is often not currently achieved~\cite{ioannidis:oxymoron}. The data collection process may have inherent biases due to contextual dependencies, temporal or other sampling biases, and simply misunderstood semantics. Inferences drawn from our personal data could be wrong, whether due to flawed algorithms, incomplete data or simply the way our attitudes and preferences change over time. User-centric controls are required, not only for consent, but for the revocation of collected personal data~\cite{whitley:informational}. In addition to a richer and more robust dialogue between regulators and the industry~\cite{forum11:_person_data}, we believe that enabling these requires stakeholders, including researchers, regulators, technologists, and industry, to establish qualitative and quantitative techniques for understanding and informing activity around human data.

Without such support it seems unlikely we will fully realise current visions of \emph{Personal APIs}~\cite{apime} that enable voluntary participation in information marketplaces~\cite{DBLP:journals/corr/abs-1205-0030,hotplanet13}: a survey of 1,464 UK consumers said that 94\% believed that they should be able to control information collected about them~\cite{bartlett:dialogue}. It is worth noting that providing such abilities might also bring benefits to data collection and processing organisations as well: the same survey reported that 65\% of respondents said that they would share additional data with organisations ``if they were open and clear about how the data would be used and if I could give or withdraw permission''.

Note that we do not suggest \emph{all} users must become continuously engaged in the collection, management and processing of their personal data. Extensive work in the context of privacy and personal data has demonstrated such features as the privacy paradox, whereby privacy only becomes a concern after a violation~\cite{barnes:privacy}, and we might reasonably anticipate that many people will not often need or desire the capacity to act within these data collection and processing systems. However, many will from time-to-time, and some enthusiasts may do so more frequently and we claim they must be supported in doing so.

Evidence suggests that mechanisms for expressing data management, such as privacy policies, are difficult both to design~\cite{trudeau:introspection} and to interpret~\cite{mcdonald:policies,leon:disclosures}, and so supporting users acting more broadly may prove a significant challenge. The interplay between data collectors and third-party data users~\cite{Krishnamurthy,MarjanTMA,Butkiewicz:webcomplexity} introduces new challenges, both to the privacy of personal data and to the understanding of this privacy: How can we accurately measure the effects of personal data collection when the effects of this collection may span multiple entities and multiple time periods? If we cannot measure these effects, then it will be hard to convince people that they should be concerned, or that they should adopt privacy mechanisms such as differential privacy~\cite{diffpriv}, privacy-preserving profiling and advertising schemes such as Adnostic~\cite{adnostic}, MobiAd~\cite{mobiad}, or PrivAd~\cite{guha.hotnets09}, or privacy metaphors~\cite{adams:med-lights,kapadia:walls} to simplify the configuration of systems.

It is also worth noting that not all activities associated with processing of personal data are harmful, and so granting users agency in these systems need not have only negative effects. Recommender systems~\cite{Ricci:2010:RSH:1941884} can provide a useful function, saving us time and effort. Live traffic updates through services such as Tom-Tom~\cite{tomtom} or Google Maps~\cite{googlemaps} assist us in avoiding traffic jams. Public health initiatives are often based on the aggregation of large quantities of highly personal data. The opportunity for data subjects to engage with data systems may enable them to correct and improve the data held and the inferences drawn, improving the overall quality and utility of the applications using our personal data.

%% file: negotiability.tex
\section{Negotiability}
\label{s:negotiability}

Having argued that personal data and its uses are complex, intimate and wide-ranging, requiring mechanisms to give people legibility and agency, to understand data and its implications and to be able to act as a result, we turn to our third theme: \emph{negotiability}. By this we mean support for people to re-evaluate their decisions as contexts change, externally (e.g.,~people and data crossing jurisdictional boundaries) and internally (e.g.,~feedback and control mechanisms have been shown to affect data sharing behaviour~\cite{feedback-chi14}).

Much current debate around use of personal data assumes data are considered a ``good'' that can be traded~\cite{acquisti:economics}, and from which economic value should be extracted~\cite{oecd:personal-data}. Although we agree it may well be possible to enable an ecosystem using economic value models for utilisation of personal data~\cite{hotplanet13,Gill:2013:BPF:2504730.2504768} and marketplaces~\cite{DBLP:journals/corr/abs-1205-0030}, we believe that power in the system is presently disproportionately in favour of the data aggregators that act as brokers and mediators for users, causing the apparent downward trajectory of economic value in the information age~\cite{Jaron}.

Effectively redressing this balance requires research to understand the \emph{contextual integrity}~\cite{nissenbaum:integrity} of uses of our personal data, and how this impacts services~\cite{shilton:stream} and new uses of our data both for research and business~\cite{mcneilly:concerns}. Contextual effects mean that data connected with people cannot realistically be considered \emph{neutral} or \emph{value-free}, leading to problems with applying concepts such as the \emph{data-driven society} or \emph{Big Data} to individuals. Expecting people to be able to self-manage their personal private data may be inappropriate given increased data collection~\cite{solove:self-management} and so legal and regulatory frameworks may need revisiting and readdressing~\cite{stodden:framework,westby:legal}. Indeed, deficiencies in the current regulatory situation around the use of \emph{big} personal data are slowly being responded to by bodies such as the EU Court of Justice~\cite{EU:data-retention} and the UK Parliament~\cite{commons:rip-offs}. But not all data are collected by third parties, and quantified self and other ``self-surveillance'' data introduce new legal challenges~\cite{kang:self-surveillance}.

Some of these issues are already being faced by researchers carrying out experiments that use personal data. Experiment design requires careful consideration of the types of data to be used and the ways in which appropriate consent to use data can be obtained~\cite{brown:consent,bps:guidelines,neuhaus:ethics,stopczynski:practitioners}. Sharing of research data sharing is becoming popular, even mandated,\footnote{\url{http://www.epsrc.ac.uk/about/standards/researchdata/Pages/policyframework.aspx}}, as a mechanism for ensuring good science and the dissemination of good science~\cite{piwowar:identifying,callaghan:output}. As a result, issues such as the privacy and ethics issues of sharing -- and \emph{not} sharing~\cite{huberman:sociology} -- data are increasingly being discussed~\cite{andrienko:dagstuhl,carusi:labyrinth,orourke:disclosure}.

Much of our presentation has focused on issues surrounding specifically personal data. The power of open data, open knowledge, and open innovation are also being widely advocated by a number of independent organisations.\footnote{e.g., \url{http://www.datasociety.net}, \url{http://okfn.org}, and \url{http://theodi.org}} The objective of these efforts is to free individuals and the Web from echo chambers and filter bubbles~\cite{pariser:filter}, and empower individuals through transparent access and audit of governments and various organisations. Examples include Open Data and public health\footnote{\url{http://openmhealth.org}}, OpenStreetMap\footnote{\url{http://openstreetmap.org}}, and open reports of crime data~\cite{police.data}. The underlying belief is that publishing data will encourage making it participatory and accessible, leading to innovation and benefit to the populace. Releasing data to the public, however, needs care and foresight into usage, correlation, and reputation side effects. Availability of certain crime data about a specific neighbourhood may in fact reinforce that area as a hub. Individuals hidden in previously anonymized, delinked personal data may become identifiable through application of newly open data sets~\cite{ohm10:broken.promis.privac}. As a result, HDI needs to take into account not only personal data, but also current \emph{and future} data sets.

Finally, as we build infrastructures and interfaces that enable users to understand and engage with data processing systems, we must consider how these will shape and be shaped by the ways that we reason about our data. The kinds of analogies we build and use in this reasoning will be informed by cultural and contextual differences and similarities and, in turn, will inform how we use, release, and distribute personal data in different communities and cultures.

%% file: conclusion.tex
\section{Conclusions}
\label{sec:conclusion}

In this paper we have presented our conception of \emph{Human-Data Interaction} (HDI) in terms of three core themes -- legibility, agency, negotiability -- in the context of our evolving data-driven society. We believe that there is a strong case for the treatment of HDI as a distinct topic due to the importance of ensuring that people remain the first consideration of a data-driven society, and the breadth of disciplines it draws on. While addressing the challenges of HDI will require a range of expertise from computer science (notably security and privacy, human-computer interaction, information systems), it also involves psychology, economics and law. Identifying it as a topic in its own right helps to ensure that it can draw on all these disciplines but is not the sole purview of a single one.

Ultimately, as recent furore over the large-scale collection and processing of data has shown (e.g.,~the revelations concerning the USA National Security Agency PRISM program, and other similar national programs; the UK National Health Service \emph{care.data} program~\cite{caredata,caredata-info}; or the UK supermarket chain Tesco's plans to scan customers' faces and sell the images to advertisers~\cite{hawkes:tesco}), public concern over these matters may initially appear negligible but can surface and grow surprisingly rapidly and with considerable force. We believe that a coherent, well-founded approach to HDI able to provide insight, tools and techniques to manage human interaction with data and data processing, is a prerequisite to building trust in such large-scale (as well as small-scale) data processing.

Specifically, we believe that technology designers must take on the challenge of building \emph{ethical systems}: technology that not only provides users with agency in intentional action but also supporting the outcomes of \emph{involuntary behaviour} being predictable by users, both as individuals and as groups. This is an important step on the path to realising \textit{calm technology}~\cite{Weiser:1997:CAC:504928.504934}.